\documentclass[a4paper,12pt]{article}
\usepackage[shortcuts]{extdash}
\usepackage{hyperref}
\usepackage[nosort]{cite}

\usepackage{color}
\usepackage{amssymb,amsmath,amsfonts,amsthm,fullpage,epic,eepic,float}

\usepackage[mathscr]{eucal}

\usepackage{rotating,indentfirst,array,varioref}
\usepackage{appendix,marginnote,tikz,pgf,mathtools}
\usepackage{setspace}

\usepackage[left=1cm,right=1cm,
    top=2cm,bottom=2cm,bindingoffset=0cm]{geometry}

\usepackage{authblk}

\usepackage{tikz,pgf}

\def\be{\begin{equation}}
\def\ee{\end{equation}}
\def\barr{\begin{array}}
\def\earr{\end{array}}

\newtheorem{Proposition}{Proposition}[section]

\newcommand{\ba}{\begin{equation}\begin{aligned}}
\newcommand{\ea}{\end{aligned}\end{equation}}

\newcommand{\bml}{\begin{multline}}
\newcommand{\eml}{\end{multline}}

\newcommand{\CC}{\mathbb{C}}

\newcommand{\ZZ}{\mathbb{Z}}
\newcommand{\PP}{\mathbb{P}}

\newcommand{\dd}{\mathrm{d}}
\newcommand{\pd}{\partial}

\newcommand{\Jac}{\mathrm{Jac}}

\newcommand{\Rc}{\mathrm{Jac}\, W}

\begin{document}

\title{GLSM for Calabi-Yau Manifolds of Berglund-Hubsch Type}

\author{Konstantin Aleshkin $^{1}$\thanks{aleshkin@math.columbia.edu},}
\author{Alexander Belavin $^{2,3,4}$\thanks{belavin@itp.ac.ru} }

\affil{$^1$ Columbia University, Department of Mathematics\\
 2990 Broadway\\ New York, NY 10027}

\affil{$^2$ L.D. Landau Institute for Theoretical Physics\\
 Akademika Semenova av. 1-A\\ Chernogolovka, 142432  Moscow region, Russia}

%\affil{$^2$ International School of Advanced Studies (SISSA),
% via Bonomea 265, 34136 Trieste, Italy}
\affil{$^3$ Moscow Institute of Physics and Technology\\
Dolgoprudnyi, 141700 Moscow region, Russia}

\affil{$^4$ Institute for Information Transmission Problems\\
 Bolshoy Karetny per. 19, build.1, Moscow 127051 Russia}

\maketitle
\abstract{In this note we briefly present the results of our computation of
  special K\"ahler geometry for polynomial deformations of Berglund-H\"ubsch type
  Calabi-Yau manifolds. We also build mirror symmetric Gauge Linear Sigma Model
  and check that its partition function computed by Supersymmetric localization  coincides with exponent of the K\"ahler potential
  of the special metric.}

\flushbottom

%%%%%%%%%%%%%%%%%%%%%%%%%%%%%%%%%%%%%%%%%%%%%%%%%%%%%%%%%%%%%%%%%%%%%%%%%%%%%%%%%%%%%%%%%%%%%%%%%%%%%%%%%%%%
%%%%%%%%%%%%%%%%%%%%%%%%%%%%%%%%%%%%%%%%%%%%%%%%%%%%%%%%%%%%%%%%%%%%%%%
%%%%%%%%%%%%%%%%%%%%%%%%%%%%%%%%%%%%%%%%%%%%%%%%%%%%%%%%%%%%%%%
%%%%%%%%%%%%%%%%%%%%%%%%%%%%%%%%%%%%%%%%%%%%%%%%%%%%%%%%%%%%%%%%%%%%%%%%%%%%%%%%%%%%%%%%%%%%%%%%%%%%%%%%%%%%

\section{Special geometry for invertible singularities}

Special K\"ahler geometry\cite{COGP} is the geometrical structure underlying particular coupling constants
in superstring compactifications (such as coupling constants of particles of vector supermultiplets
in type IIA/B compactifications). Knowledge of this metric is important for phenomenological
questions arising in superstring theories.

In a series of papers \cite{AKBA, AKBA2, AKBA3,AKBA4} we showed that geometry of complex
moduli of Calabi-Yau manifolds associated with Landau-Ginzburg models can be effectively computed
through the computations in the corresponding Landau-Ginzburg model. These results were
used to confirm the Refined Swampland Finite Distance conjecture \cite{Katmadas:2017gxn,Blumenhagen:2018nts,Blumenhagen}.

The special K\"ahler metric (Weil-Peterson metric) has a K\"ahler potential whose logarithm is a Hermitian pairing of
the period integrals of the holomorphic volume form on the Calabi-Yau manifold. It is much easier
to compute these data on the Landau-Ginzburg side.

\paragraph{Invertible singularities}

A polynomial $W(x_1, \ldots, x_N)$ in $N$ variables (Landau-Ginzburg superpotential) is called an
invertible singularity if it is of the form
\begin{equation}
  W_0(x_1, \ldots, x_N) = \sum_{i=1}^N\prod_{j=1}^N x_j^{M_{ij}}
\end{equation}
for an invertible matrix $M$ with positive integer coefficients. Such a polynomial is always weighted
homogeneous:
\begin{equation}
  W_0(\lambda^{q_1} x_1, \ldots, \lambda^{q_N} x_N) = \lambda W_0(x),
\end{equation}
where $q_i = \sum_{j} M^{-1}_{ji}$ are positive rational numbers. Let $d$ be the least common denominator
of the integers $k_1, \ldots, k_N$ so that $k_i = d q_i$.

Then the equation $W_0(x) = 0$ is well-defined in the weighted projective space
\begin{equation} \label{wps}
  \PP^{N-1}_{(k_1, \ldots, k_N)} = (\CC^N - \{0\})/\CC^*,
\end{equation}
where $\CC^*$ acts on coordinates of $\CC^N$ with weights $k_1, \ldots, k_N$ correspondingly.
Typically these weighted projective spaces are smooth orbifolds. Zero locus $X_0$ of the superpotential
$W_0(x)$ is an orbifold as well. In order for $X_0$ to be smooth orbifold (or quasi-smooth variety)
the superpotential must be transverse: $\dd W_0(x) = 0 \equiv x = 0$. By classification of singularities
performed by Kreuzer and Skarke \cite{Kreuzer:1992bi} $W_0(x)$ must be a sum of three basic or
atomic types:
\begin{equation}
  \begin{aligned} \label{blocks}
    x^A &- \text{ point},\\
    x_1^{A_1} x_2 + x_2^{A_2} x_3 + \cdots + x_n^{A_n} &- \text{ chain},\\
    x_1^{A_1} x_2 + x_2^{A_2} x_3 + \cdots + x_n^{A_n} x_1 &- \text{ loop},
  \end{aligned}
\end{equation}
where each variable belongs to only one of the blocks.

If $d = \sum_{i \le N} k_i$  then the quasi-smooth variety $X_0$ is actually a Calabi-Yau variety.

In this note we describe the geometry of the (formal neighbourhood of the) space of polynomial deformations
of $W_0(x)$ which describes a certain subspace in the space of deformations of complex structures on the
Calabi-Yau manifold $X_0$.

We note that invertible singularities were studied in the context of Berglund-H\"ubsch-Krawitz mirror symmetry
\cite{BerHub, Krawitz}. 

\paragraph{The space of states and cohomology}
Given an isolated (even transverse) singularity $W_0(x)$ its Jacobi or Milnor ring is a space of infinitesimal
deformations modulo infinitesimal coordinate changes
\begin{equation}
  \Jac(W_0) = \frac{\CC[x_1, \ldots, x_N}{\langle \pd_1 W_0, \ldots, \pd_N W_0 \rangle}.
\end{equation}
Let us denote $\{e_1, \ldots, e_h\}$ to be a set of homogeneous degree d monomials which span a basis
of the degree d part of $\Jac(W_0)$, where we denote $e_l = x_1^{S_{l1}} \cdots x_N^{S_{lN}}$ and
$l=1,...,h$. 
The general  polynomial weighted homogeneous deformation of $W_0(x)$ can be expressed as
\begin{equation}
  W(x, \phi) = W_0(x)  + \sum_{l=1}^h {\phi_l} e_l.
\end{equation}
This family is a particular deformation of complex structures on $X_{\phi} = \{W(x, \phi) = 0\}$ and each monomial
$e_l$ corresponds to a cohomology class $e_l\to H_{N-3,1}(X_{\phi})$.

The Jacobi ring of an invertible singularity is a tensor product of the Jacobi rings of its atomic blocks.
The latter ones are described by the following proposition \cite{Kreuzer:1992bi}:

\begin{Proposition} \label{i:basis}
  Milnor rings for three elementary blocks (atomic types) have the following bases:
  \begin{enumerate}
  \item Point, $W(x) = x^A$,
    \begin{equation}
      \Rc_0 = \langle x^a \rangle_{a=0}^{A-2}
    \end{equation} 
  \item Chain, $x_1^{A_1} x_2 + x_2^{A_2} x_3 + \cdots + x_N^{A_N}$. 
    Then the basis for $\Rc_0$ is most conveniently written recursively:
    \begin{equation}
      \Rc_0 = \bigoplus \langle x_{1}^{a_1} \cdots x_N^{a_N} \rangle,
    \end{equation}
    where 1) either $a_1 \le A_1-2$ and $a_i \le A_i - 1$ for $i>1$ or 2)
    $a_1=A_1-1, \, a_2=0$ and $x_3^{a_3} \cdots x_N^{a_N}$ define a basis element
    of the chain $x_3^{A_3} x_4 + x_4^{A_4} x_5 + \cdots + x_N^{A_N}$.
    
  \item Loop, $x_1^{A_1} x_2 + x_2^{A_2} x_3 + \cdots + x_N^{A_N} x_1$
    \begin{equation}
      \Rc_0 = \bigoplus_{k_i=0}^{A_i-1}\langle x_{1}^{a_1} \cdots x_N^{a_N} \rangle,
    \end{equation}
    where $0\leq a_{i}<A_{i}$.
  \end{enumerate}
\end{Proposition}

The $\ZZ_d \subset \CC^*$ from the definition of the weighted projective space~\eqref{wps} acts on $\CC^N$ and on the
Jacobi ring $\Rc_0$. The invariant part $\Jac(W_0)^{\ZZ^d}$ decomposes into $\Jac(W_0)^{\ZZ^d} = \bigoplus_{k=0}^{N-2} \Jac(W_0)^{dk}$,
where $\Jac(W_0)^{dk}$ consists of elements of weight $dk$. In what follows we specify to the case $N=5$. In this case
\begin{equation}
  \Jac(W_0)^{\ZZ^d} = \langle 1 \rangle \oplus \Jac(W_0)^{d} \oplus \Jac(W_0)^{2d} \oplus \langle \mathrm{Hess} W_0  \rangle
\end{equation}
and $\Jac(W_0)^{2d} \simeq \Jac(W_0)^d$ and has a natural dual basis. We denote the monomial basis
of $\Jac(W_0)^{\ZZ^d}$ by $\{e_a \}_{a=0}^{1+2h}$, where $e_a = \prod_{j=1}^5 x_j^{a_j}$.

\paragraph{Period integrals}

The period integrals of the holomorphic form are
\begin{equation}
  \sigma[\gamma_a](\phi) := \int_{\gamma_a} \Omega_{\phi},
\end{equation}
where $\gamma_a \in H_3(X_0)$ is a  cycle dual to the  $e_a$ - a basis element of the invariant ring. (For more  information on  defining such cycles, see \cite{AKBA3}).
$\Omega_{\phi}$ is a holomorphic volume form on $X_{\phi}$ for each
value of $\phi$. It is well-known that they satisfy the Picard-Fuchs partial differential equations and
in principle can be computed using the Frobenius method. For the considered case Frobenius-type basis of the periods can be computed following our approach in ~\cite{AKBA, AKBA2,AKBA3, AKBA4} as a power
series in $\phi_1, \ldots, \phi_{h}$ around zero:
\begin{equation} \label{periods}
  \sigma_a(\phi) =  \sum_{\sum m_l S_{li} = 
	a_i\, \textrm{mod}\, n_{j} M_{ji}} \prod_{s=1}^h  \frac{\phi_s^{m_s}}{m_s!} 
\prod_{i=1}^5  \frac {\Gamma(\sum_j(\sum_l m_l S_{lj} +1) B_{ji})}{\Gamma(\sum_j(\mu_j +1) B_{ji})},
\end{equation}
where $B_{ji}=(M^{-1})_{ji}$ and $n_{j}$ are some non-negative integers.\\ 
%%%%%%%%%%%%%%%%%%%%%%%%%%%%%%%%%%%
The expresssion \eqref{periods} can be rewritten as 
\begin{equation} \label{periods2}
  \sigma_a(\phi) = \sum_{n_1, \ldots, n_5} \prod_{i=1}^5\left((a_j+1) B_{ji} \right)_{n_i} 
\sum_{\sum_l m_l S_{li} = a_i+n_j M_{ji}} \prod_{s=1}^h
  \frac{\phi_s^{m_s}}{m_s!}, 
\end{equation}
where the raising factorials are $(x)_n = \Gamma(x+n)/\Gamma(x)$.

\paragraph{The formula for the special K\"ahler potential}

The exponent of the K\"ahler potential of the Weil-Peterson metric on the deformation space is a Hermitian expression in the periods.
The $W_0(x)$ has a huge discrete group of diagonal symmetries of which $\ZZ_d$ is a subgroup. The diagonal symmetry group naturally
acts on the deformation space with coordinates $\{\phi_s\}_{s=1}^h$. For most invertible singularities the period integrals~\eqref{periods} 
form one-dimensional representations of the diagonal symmetries group with different weights. This implies that the Hermitian pairing should be diagonal in $|\sigma_a(\phi)|$. Indeed, we show that this is the case for all invertible singularities and the coefficients are products of ratios of Gamma-functions in the normalization~\eqref{periods}:
\begin{equation} \label{mainFormula}
  e^{-K(\phi, \bar{\phi})} = \sum_{a=0,..., 2d-1} (-1)^{|a|/d} \prod^\prime_{i \le 5}\frac{\Gamma\left((a_j+1)B_{ji}\right)}{\Gamma\left(1-(a_j+1)B_{ji}\right)}
  |\sigma_a(\phi)|^2,
\end{equation}
where $|a| = \sum_j k_ja_j$ is the weight of the monomial $e_a$ and $\prod^{\prime}$ denotes product over all terms where arguments of Gamma functions
are non-integer.

\section{Localization and Mirror symmetry}

Gauge Linear Sigma Models (GLSM) \cite{Witten:1993yc} have supersymmetric backgrounds on the round sphere $S^2$.
Corresponding partition function was computed in~\cite{BeniniCremonesi, Doroud:2012xw}.\\
%%%%%%%%%%%%%%%%%%%%%%%%%%%%%%%%%%%%%%%%%%%%%%%%%%5
For the case when the gauge group of the model  $G=\prod_{l=1}^{h}U(1)_l$,  the  chiral matter fields $\Phi_a,a=1,...,N=h+5$ have $R$-charges $q_a$  and their  charges with respect to the gauge group $U(1)_l$ are denoted by $Q_{la}$ the partition function looks as
\begin{equation}\label{Benini}
   Z_{Y}=\sum_{\bar{m} \in \Lambda}
\prod_{l=1}^{h}\int_{\mathcal{C_l}}\frac{d\tau_{l}}{(2\pi i)}
\left(z_{l}^{-(\tau_l - \frac{m_l}2)}\bar{z}_{l}^{-(\tau_l + \frac{m_l}{2})}\right) 
 \prod_{a=1}^{h+5}\frac{\Gamma\bigl(\frac{q_a}{2}+\sum_{l=1}^{h}Q_{la}(\tau_l - \frac{m_l}2)\bigr)}
{\Gamma\bigl(1-\frac{q_a}{2}-\sum_{l=1}^{h}Q_{la}(\tau_l + \frac{m_l}2)\bigr)}
\end{equation}
where  $z_{l}=e^{-(2\pi r_{l}+i\theta_{l})}$, $r_l, \theta_l$ 
are Fayet-Iliopoulos parameters  and theta angles for our GLSM. 
The summation by $\bar{m} = \{m_1, \ldots, m_h\}$ goes over the set specified below
by~\eqref{quantization}.\\
%%%%%%%%%%%%%%%%%%%%%%%%%%%%%%%%%%%%%%%%%%%%%%
The JKLMR conjecture \cite{Jockers:2012dk} claims that the partition function computes the exponent of K\"ahler potential of the quantum corrected metric on the K\"ahler moduli space of GLSM in the Calabi-Yau case.

Mirror symmetry connects quantum corrected K\"ahler moduli space of one model with
complex structures moduli space of another. For all Calabi-Yau threefolds given by invertible
singularities we construct  the corresponding  mirror symmetric GLSM and compute their partition functions following the method suggested in \cite{Aleshkin:2018pfc,Aleshkin:2018tbx}.
Then we  explicitly check that they coincide with exponents of K\"ahler potentials of Weil-Petersson
metrics on the deformation spaces of the invertible singularities.  

To do this, we use Batyrev's approach to the mirror symmetry \cite{Batyrev:1994hm} in the same way as in  \cite{Aleshkin:2018pfc,Aleshkin:2018tbx,Eremin}   to construct GLSM for  Fermat hypersurface cases.\\
%%%%%%%%%%%%%%%%%%%%%%%%%%%%%%%%%%%%%%%%%%%%%%%
The idea is as follows. Let Calabi-Yau threefold $X$ be  defined as a hypersurface in a weighted projective space $\mathbb{P}^4_{(k_1, \dots, k_5)}$ and  given by zero locus of the polynomial $W(x,\phi)$. Exponents of the monomials that make up the polynomial  $W(x,\phi)$ determine the finite set $\Vec{V}_a \ a=1,\dots,N$ and, thus, define Batyrev's polytope  $\Delta_X$ \cite{Batyrev:1994hm}.

 Knowing  the set of vectors $\Vec{V}_a$ we can construct a fan \cite{Mirror}, which defines another toric variety. Then  Calabi-Yau manifold $Y$, the mirror to $X$, is defined as a hypersurface in this  variety given by the zero locus of  the  homogeneous polynomial $W_Y$. Using this fan we also  find  the corresponding GLSM with its gauge group $G=\prod_{l=1}^{h}U(1)_l$ and the set of  charges $Q_{la}$ of its chiral multiplets.  The charges appear as a coefficients of the  linear relations between the vectors of the fan and set the weights of the toric variety

%%%%%%%%%%%%%%%%%%%%%%%%%%%%%%%%%%%%%%%%%
In the considered case we have the following expression for the deformed invertible singularity:
\begin{equation}
  W(x,\phi) = \sum_{i=1}^5\prod_{j=1}^5 x_j^{M_{ij}} + \sum_{l=1}^h \phi_l \prod_{j=1}^5 x_j^{S_{lj}},
\end{equation}
which defines the Calabi-Yay family $X$. The  set of the vectors  $\Vec{V}_a \ a=1,\dots,N$ in this case is
 \begin{equation}V_{ai}=
 \begin{cases}
 5\delta_{a,i},\quad 1\le a\le 5,\\
 S_{a-5,i},\quad 6\le a\le h.
 \end{cases}
 \end{equation}

The corresponding   GLSM has   the gauge group  $G=\prod_{l=1}^{h}U(1)_l$, 
the vector superfields $\{V_l\}$, $l=1,...,h$,  and $h+5$ chiral superfields $\{\Phi_a\}$, $a=1,...,h+5$, interacting with  $\{V_l\}$ by  integer charges ${Q}_{la}$. 
It is convenient \cite{Aleshkin:2018tbx} to introduce instead of the  integer charges ${Q}_{la}$ another  ``rational charge matrix'' ${Q'}_{la}$
which satisfies
\begin{equation}
\sum_{a \le 106} Q'_{al} \Vec{V}_a = 0,
\end{equation}
which looks as
\begin{equation} \label{l:QInvertible}
  Q'_{la} = \begin{cases}
    S_{lj} B_{ja} ,\quad 1\le a\le 5,\\
    -\delta_{a-5, l}, \quad a> 5.
  \end{cases}
\end{equation}
${Q'}_{la}$ is connected with the charge matrix  ${Q}_{la}$, whose entries are integers, 
by a rational invertible $h \times h$ matrix.
In terms of the new  rational basis of charges the integrality condition  for $m_l$  modifies as follows:
\begin{equation} \label{quantization}
  m_l \in \ZZ, \; \to \bar{m} _l Q'_{la} \in \ZZ.
\end{equation}

 Then the generic  expression \eqref{Benini} for the partition function    specifies in our case to
\begin{equation}\label{Benini1}
   Z_{Y}=\sum_{\bar{m} \in \Lambda}
 \int\limits_{\mathcal{C}_{l}}\frac{d^h\tau}{(2\pi i)^h}
\frac{1}{z_{l}^{(\tau_l - \frac{m_l}2)}\bar{z}_{l}^{(\tau_l + \frac{m_l}{2})}}
 \prod_{l=1}^{h}\frac{\Gamma\bigl(\frac{q_{l+5}}{2}-(\tau_l - \frac{m_l}2)\bigr)}
{\Gamma\bigl(1-\frac{q_{l+5}}{2}+(\tau_l + \frac{m_l}2)\bigr)}
\prod_{i=1}^{5}\frac{\Gamma\bigl(\frac{q_i}{2}+\sum(\tau_l - \frac{m_l}2)S_{lj}B_{ji}\bigr)}
{\Gamma\bigl(1-\frac{q_i}{2}-\sum(\tau_l + \frac{m_l}2)S_{lj}B_{ji}\bigr)},
\end{equation}
where the summation over $\bar{m}={m_1,...,m_h}$ goes over the  specified by~ \eqref{quantization}.\\
We also have to assign the R-charges to the chiral fields in appropriate way \cite{Witten:1993yc}.
We can choose them be zero besides one $q_h=2$.\\
After this setting we compute the integral for $|z_l|>1$  closing the integration  contours to the right half-planes.
The result reproduces the formula~\eqref{mainFormula} in the
Landau-Ginzburg phase up to the simple coordinate change between Fayet-Iliopoulos parameters of the GLSM
and the complex deformation parameters $\{\phi_l\}_{l=1}^h$ which looks as:
\begin{equation}
    z_l = -\phi_l^{-1}, \;\;\; 1\ \le l \le h,
\end{equation}
which is the mirror map. This gives the realization of  the mirror version of the JKLMR conjecture 
 \cite{Jockers:2012dk} for the considered class of Calabi-Yau families.

\section*{Conclusion}
Thus,  starting from the model with of Berglund-H\"ubsch type   Calabi-Yau manifolds $X$ we have constructed the  $\mathcal{N}=(2,2)$ Gauged Linear Sigma Model with the manifold of supersymmetric vacua $Y$, which is the mirror for $X$.  
Having computed  Special geometry on the moduli space of complex structures on $X$ and using Batyrev's approach \cite{Batyrev:1994hm} to Mirror symmetry we have checked the JKLMR conjecture \cite{Jockers:2012dk}. A detailed description of the procedure for obtaining these results will be the subject of another article.

\section*{Acknowledgments}

We are grateful to N.~Doroud, A.~Litvinov and M.~Romo  for the useful discussions. This work  has been supported by the Russian Science Foundation under the grant 18-12-00439.

\end{document}